\documentclass[sigconf]{acmart}

\usepackage{booktabs} 
\usepackage{enumitem}
\usepackage{subcaption}
\usepackage[utf8]{inputenc}
\setcopyright{rightsretained}
\usepackage{xcolor}

\acmDOI{}

\acmISBN{}

\acmConference[arXiv]{arXiv}{2017}{Kevin Jasberg and Sergej Sizov} 
\acmYear{2017}
\copyrightyear{2017}

\acmPrice{15.00}

\begin{document}
\title{Bayesian Brain meets Bayesian Recommender\\ Towards Systems with Empathy for the Human Nature}

\author{Kevin Jasberg}
\affiliation{%
  \institution{Web Science Group\\Heinrich-Heine-University Duesseldorf}
  \city{Duesseldorf} 
  \state{Germany} 
  \postcode{45225}
}
\email{kevin.jasberg@uni-duesseldorf.de}

\author{Sergej Sizov}
\affiliation{%
  \institution{Web Science Group\\Heinrich-Heine-University Duesseldorf}
  \city{Duesseldorf} 
  \state{Germany} 
  \postcode{45225}
}
\email{sizov@hhu.de}

\renewcommand{\shortauthors}{}

\begin{abstract}
In this paper we consider the modern theory of the Bayesian brain from cognitive neurosciences in the light of  recommender systems and expose potentials for our community. 
In particular, we elaborate on noisy user feedback and the thus resulting multicomponent user models, which have indeed a biological origin. In real user experiments we observe the impact of both factors directly in a repeated rating task along with recommendation. 
As a consequence, this contribution supports the plausibility of contemporary theories of mind in the context of recommender systems and can be understood as a solicitation to integrate ideas of cognitive neurosciences into our systems in order to further improve the prediction of human behaviour.
\end{abstract}

%
%


\maketitle

\section{Introduction}
In our community of recommender systems, there are continual efforts to make predictions more precise and systems more efficient and user-friendly.
In doing so, the classic approach is to model the relationship between a user and items in terms of optimising a target function in order to predict future user decisions based on training data.
However, there are two major problems that are caused by human nature. First of all, many studies prove that users are not entirely certain about a decision so that a given rating may fluctuate when the rating task is repeated (noisy user feedback) \cite{Hill, UMAP, RateAgain, LikeLikeNot}.
Second, optimising a target function might not sufficiently account for dynamic changes in behaviour (need for multicomponent user models) which again can be proven in systematic user experiments  \cite{Sizov}. 
The theories of cognitive neurosciences know these phenomena and describe these aspects of human cognition by means of probabilistic models. For example, the origin of volatile decision-making and noisy user feedback is due to the irregular transmission of informations through the synaptic cleft. A Bayesian formulation of this effect leads directly to multicomponent models as an explicit consequence of user noise. Both of these factors, noisy user feedback and multicomponent models, have recently been proven to have a significant impact on prediction quality in recommender systems \cite{LikeLikeNot, MagicBarrier, cub}.
Therefore, this contributions seeks the benefits of implementing neuroscientific models and will give experimental indication for possible deficiencies within the field of recommender research in case of omission.

\section{Theory and Experiments} 
\begin{figure}[t]
\includegraphics[width=.99\linewidth]{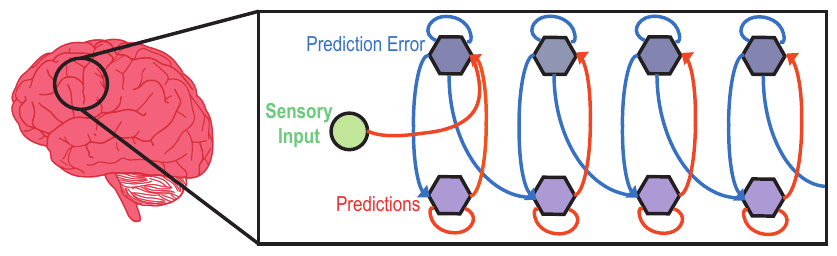}
\caption{\small Hierarchical message passing through cortex layers.\vspace*{-4ex}}
\label{HMODEL}
\end{figure}

The Bayesian brain theory is a composition of Bayesian inference and theoretical neurology.
We will first demonstrate similarity to Bayesian recommender systems and then discuss the modelling of noisy data as well as multicomponent user models.

\subsection*{Bayesian Learning Basics}
From the perspective of cognitive neuroscience, it all starts with the brain observing sensory input $Y$ (visual, auditory, etc.) and making an estimate of the state of the world $X$ \cite{BBTbook}.
To continiously improve this estimation - or subjective beliefs to be more precisely -  the brain has to learn by comparing reality against predictions based on these beliefs \cite{FristonNature}. This makes the human brain a highly sophisticated recommender system itself. The Bayes Theorem provides the basis for the processing of beliefs along with real world evidence. Those confirmed or modified beliefs are thereby brought to ever new situations. Mathematically spoken, the posterior serves as prior in subsequent cognitions. For multiple independent sensory observations $y=(y_1,\ldots,y_n)$ we yield  
\begin{equation}
P(X\vert y)\propto P(X\vert y_1)\cdot\ldots\cdot P(X\vert y_{t-1})\cdot P(X\vert y_t)
\end{equation}
where $\prod_{i=1}^{t-1}P(X\vert y_i)$ is the posterior of $X$ given sensory data until time $t-1$ and serves as prior for time $t$ (learning from the past). When the world state itself changes while making observations, we need to consider a transition probabilities $P(X_t\vert X_{t-1})$
\begin{equation}
P(X\vert y)\propto 
P(X_t\vert X_{t-1}) \cdot \prod_{i=1}^{t-1} P(X_{t-1}\vert y_{i}) \cdot P(X\vert y_t) 
\end{equation}
which is the basis of learning by iteration. Here, we can already see the similarity to Bayesian recommender systems clearly \cite{MachineLearning}. But how does the brain actually model prior probabilities? Mathematically, this question is pointless when hierarchical networks are used since these models optimise the priors themselves through mutual back-propagation of predictions and forward-feeding of prediction errors \cite{FristonNature} as to see in Figure \ref{HMODEL}. The optimisation task itself is done by neuron clusters (agents) via minimising the so called free energy \cite{FristonNature}. In consequence, beliefs do exist in the form of probability densities, from which particular draws are made for decision making. We will see later that these distributions indeed exist in the case of product ratings.

\subsection*{Neural Noise and Decision-Making}

\begin{figure}[t]
\includegraphics[width=.65\linewidth]{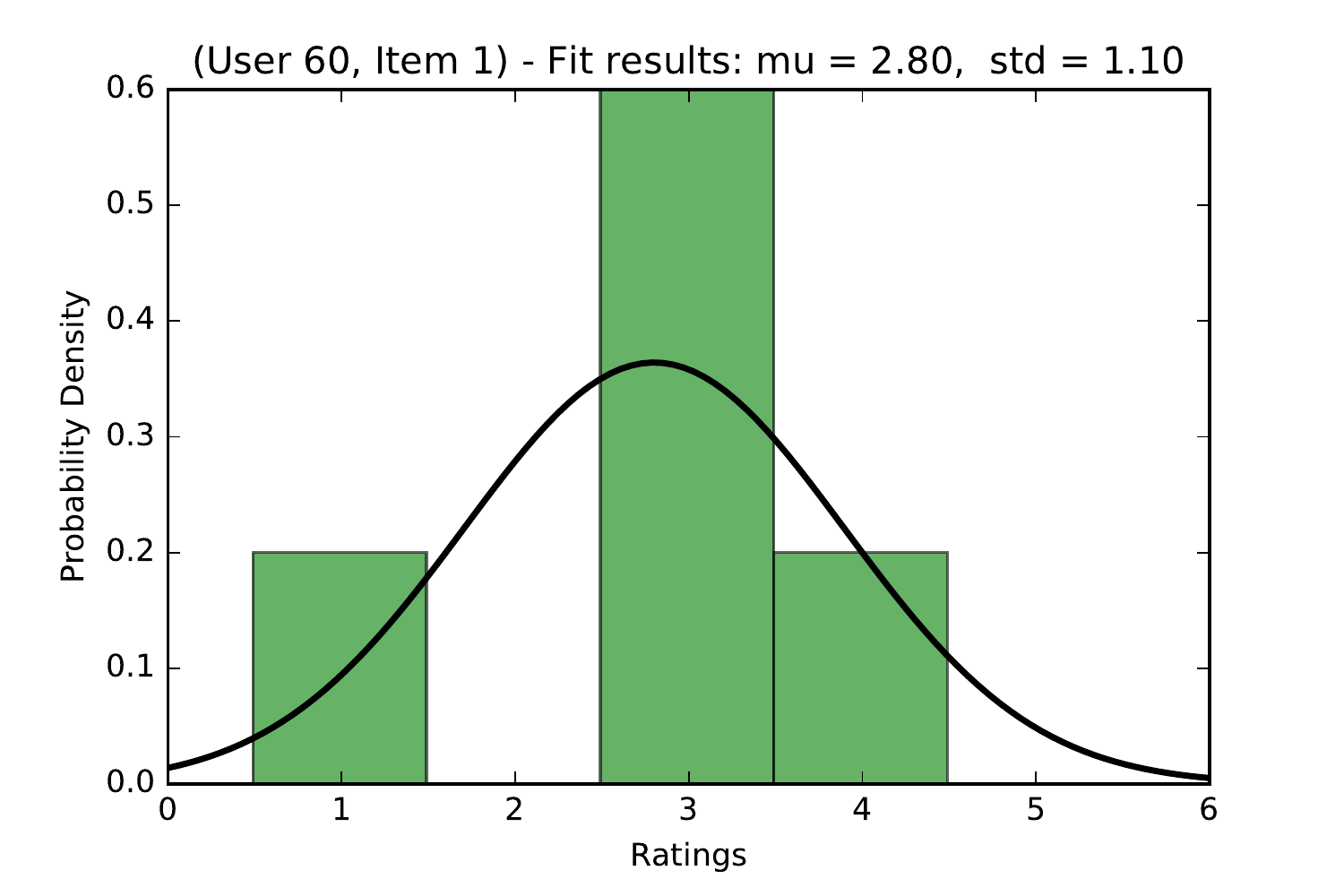}
\caption{\small Example of variable responses to the same item\vspace*{-4ex}}
\label{UserResponses}
\end{figure}

Message passing works by relaying electrical signals from one neuron to the next through the synaptic gap by means of neurotransmitters. However, same signals never result into emitting the same amount of transmitters (neural noise) \cite{FristonNature, Friston, Noise1,Noise2}. This noise may raise too weak (or sufficient) signals above (or below) a certain threshold, causing a neuron to inhibit (or to fire) \cite{BBTbook}. In fact, for the firing of a neuron, we can only specify a probability \cite{BBTbook, Gerstner}.
In a recommender's language, biological irregularity implies that every time a decision-making is repeated, other prior probabilities might be used and thus the resulting belief is never quite the same as before.
In a systematic experiment, real users repeatedly rated theatrical trailers on a 5-star scale. It turns out that only 35\% of all users show constant rating behaviour, whereas about 50\% use two different answer categories and 15\% of all users make use of three or more categories. Figure \ref{UserResponses} is a characteristic example for these results. This sows that individuals are not able to perfectly reproduce their decision-making. 
These fluctuations can be explained by the theory of neural noise, and have a direct effect on recommender systems.
Assuming the model based prediction to be $\pi=3$, a user rating $r=4$ can not be seen as a deviation according to Figure \ref{UserResponses}.
Furthermore, by gathering information about temporary belief posteriors \cite{vonmirgeklaut}, it can be proven that all user responses (aggregated for each item) hold the same expectation. This is an indication for a common source of noise with constant magnitude, i.e. the manifestation of neural noise.   

\subsection*{Modelling User Preferences}

When it comes to a repeated product rating where the participant does not remember his previous response, as induced in our controlled experiment, the process of decision making is restarted. Accordingly, the participant receives a new and slightly different belief distribution for each rating trial. This has been mathematically explained in \cite{Friston} and implies the need for multicomponent user models, which have recently been adressed in recommender systems research \cite{cub, Sizov}.
\begin{figure}[t]
\includegraphics[width=.73\linewidth]{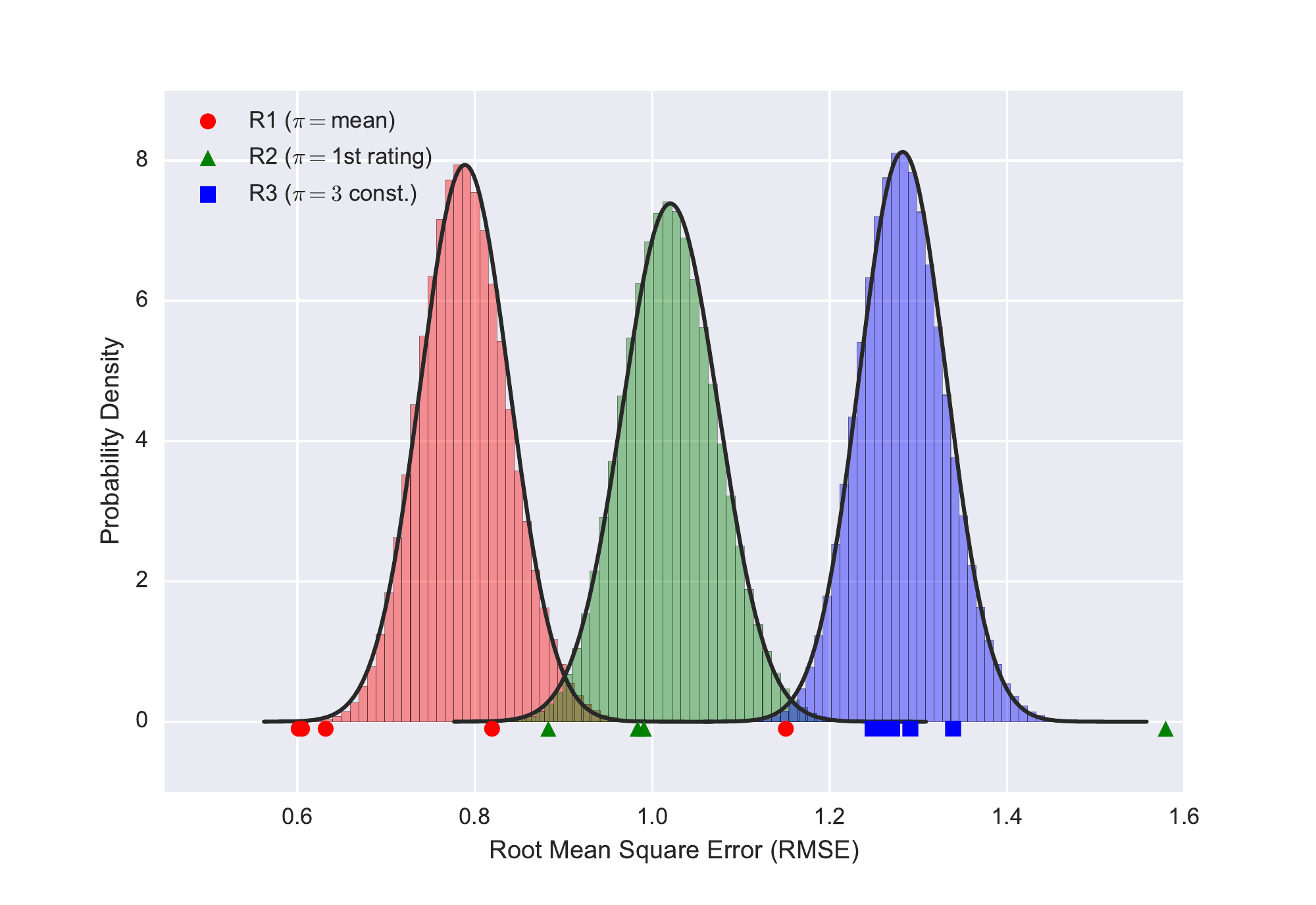}\vspace*{-4ex}
\caption{\small Precision of 3 recommender systems\vspace*{-4ex}}
\label{RMSE}
\end{figure}
Figure \ref{RMSE} shows the RMSEs of three different systems utilised in our experiment, based on one-component models along with the scores each system has achieved in each trial. It is apparent that some draws (scores) can not be drawn from the corresponding distribution. This indicates that users had changed their rating behaviour and sampled from different distributions for different trials. It can be proven via hypothesis testing that rating behaviour of trial 1 and 5 significantly deviates from trials 2 to 4. 
This can be explained by memory effects: In trial 1, decision-making was initialised for the first time.
After the trial 2, participants got aware that the experiment was about repetition and so started to remember their ratings. Therefore, belief distributions remained more or less the same. After the trial 4 plus constant addition of new distractors, short-time memory was not able to keep all previous information and further decision-making produced different belief distributions again. These findings within simple recommendation scenarios can be entirely described by the Bayesian brain theory and may help systems to learn human behaviour more naturally.

%
%
%

\section{Conclusion}

We have shown that the Bayesian brain theory uses the same models as Bayesian recommender systems.
In addition, we have shown that the Bayesian brain theory is already in a position to model quantities whose impact can be  seen in recommendation scenarios.
We have demonstrated this by the example of noisy user feedback and multicomponent models.
The impact of these factors has been discussed briefly, i.e. the interpretation of correct and false predictions has to be considered more differentiated whereas the choice of user models holds a strong dependency on time.
We therefore recommend to adopt corresponding models from neurosciences in order to optimise recommender systems in terms of imitating human behaviour.

\bibliographystyle{ACM-Reference-Format}
\bibliography{Literatur} 

\end{document}